# Quasi-diurnal Kelvin Wave Observed in the Atmosphere of Mars by the Pressure Sensor on the InSight Lander

**Anzu Asumi[1], Jorge Hernández-Bernal[2], Kaoru Sato[1], and Aymeric Spiga[2]**

[1]Department of Earth and Planetary Science, Graduate School of Science, The University of Tokyo, Tokyo 1130033, Japan

[2]Laboratoire de Météorologie Dynamique, Sorbonne Université, Paris 75005, France

Corresponding author: Anzu Asumi (asumi@eps.s.u-tokyo.ac.jp)

**Key Points:**

- Quasi-diurnal Kelvin wave was detected in InSight surface pressure timeseries using singular spectrum analysis.
- During the early stage of the dust event C in MY34, Kelvin wave and diurnal tide were transiently amplified when they are aligned in phase.
- Mars Planetary Climate Model realistically reproduces the characteristics of surface pressure as observed by InSight.

## Abstract

During the initiation phase of Martian dust storms, a spectral signature of a zonal wavenumber-1 normal mode Kelvin wave (K1) has been observed in several surface pressure records. However, the K1 signal has not been extracted directly since its eigenfrequency is close to that of the diurnal tide, making their separation difficult. We apply singular spectrum analysis to more than one Martian year of InSight surface pressure observations, including the MY34 C dust storm (C34). This analysis enables us to directly extract the transient K1 signal. Clear K1 signals appear during the Northern-Hemisphere spring and autumn with the average amplitudes of ∼1–2 Pa. They reach ∼8 Pa during the C34 storm, which is substantial compared with a simultaneous diurnal signal (S1) maximum of ∼39 Pa. Near the peak time, K1 and S1 are in phase, and unexpectedly the S1 period shortens to ∼23.5 Mars hours. Subsequently, K1 rapidly decays and the S1 period recovers to ∼24 Mars hours. The phase relation between K1 and S1 as well as the S1 period shortening indicates that a resonance occurred between S1 and K1. Comparison with simulations at the model grid point closest to the InSight landing site shows that the Mars Planetary Climate Model realistically reproduces the observed characteristics of the surface pressure variations.

## Plain Language Summary

Mars' atmosphere can be ringing by large-scale waves that travel around the planet, much like vibrations in a musical instrument. One of these waves, called the zonal-wavenumber-1 Kelvin mode, has a period close to one Martian day. Because its period is so similar to the daily atmospheric oscillations driven by solar heating, the two signals are difficult to separate in observations. We analyzed surface-pressure measurements at quite short time intervals by the InSight lander for more than one Martian year. Using a data-driven method allowed us to identify the Kelvin-mode signal directly. The Kelvin-mode signal was usually weak, but it became much stronger during a regional dust storm in Mars Year 34. At the same time, the Kelvin mode and the daily tidal signal became nearly synchronized, and the apparent period of the daily tide temporarily shortened. After the Kelvin mode weakened, the daily tide returned to its usual period. These results provide evidence that a naturally occurring atmospheric wave interacted with the daily thermal tide during the dust storm. Such interactions may help explain rapid changes in Martian atmospheric circulation during dusty periods.

## 1 Introduction

Atmospheric normal modes are free oscillations theoretically obtained by solving the three-dimensional primitive equations linearized about a motionless basic state whose temperature depends only on height (e.g., Diky and Golitsyn, 1968; Sakazaki & Hamilton, 2020). In contrast, atmospheric thermal tides are periodically forced oscillations driven by solar heating. Whereas the periods of the forced oscillations are fixed at one sol and its harmonics, the eigenfrequencies of the free oscillations vary depending on the vertical structures of atmospheric temperature. The foundational theoretical framework for atmospheric tides and normal modes was established by Chapman and Lindzen (1969).

In the Martian atmosphere, Hamilton and Garcia (1986) theoretically determined the frequencies of short-period free oscillations in the Martian atmosphere. One of these is the eastward-propagating Kelvin mode with zonal wavenumber 1, hereafter referred to as K1. For a 200 K isothermal carbon-dioxide atmosphere, they derived a theoretical period of K1, which is approximately 23.2 Mars hours with an equivalent depth of 13.7 km. Because the eigenfrequency of K1 is sensitive to the global-mean atmospheric temperature, its period is expected to vary seasonally. During the cold aphelion season in Northern Hemisphere summer, the K1 period approaches one sol, whereas at higher temperatures, it shifts away from the diurnal period. During Northern Hemisphere summer, K1 can therefore resonate with a diurnal eastward nonmigrating tide (DE1), which has nearly the same eastward propagation and zonal wavenumber (Wilson and Hamilton, 1996). Indeed, several studies using satellite temperature observations have shown that, during NH summer, the amplitudes of the diurnal eastward nonmigrating tide (DE1) are largest, suggesting the resonant enhancement of DE1 with K1 (e.g., Hinson et al., 2008; Guzewich et al. 2012). Although numerous studies have examined DE1 (e.g., Forbes and Hagan, 2000; Withers et al., 2003), little attention has been given to K1 itself. In addition, the relation between the diurnal tide and K1 during periods other than the NH summer—such as the dusty season—remains largely unclear.

Previous studies reported the relation between K1 and dust events on Mars for both episodic global dust storms and annually-recurring regional dust storms. Tillman (1988) examined surface pressure data obtained from the Viking landers by applying the maximum entropy method (MEM) for spectral analysis. They showed that K1 was present during the initiation phase of the 1977A, 1977B, and 1982A global dust storms, and suggested that K1 may play a role in triggering them. An initial analysis of surface pressure data from the Rover Environmental Monitoring Station (REMS) on the Curiosity Rover in Gale Crater (e.g., Harii et al., 2013) revealed an abrupt phase shift in the diurnal tide during the expansion of a regional dust storm, which was interpreted as resulting from interference between DE1 and DW1 (Haberle et al., 2014). Zurita-Zurita et al. (2022) analyzed the same data. They decomposed the time series of surface pressure into its Fourier components and interpreted the observed abrupt sol-to-sol phase change of the diurnal tide as signatures of K1. Abrupt phase changes in the diurnal tide were observed prior to the Z, A, and C storms. These annually recurrent storms occur every year in absence of a global dust storm: the Z storm occurs during the pre-equinox period in $L_s \sim 120°–160°$; the A storm occurs during the post-equinox period in $L_s \sim 190°–240°$; and the C storm occurs during the post-solstice period in $L_s \sim 300°–335°$ (Kass et al., 2016). These findings suggest that K1 is enhanced during the initiation of dust storms.

To go further in studying K1 and its impact on weather and climate on Mars, direct detection of this mode is needed. The K1 direct detection is generally difficult because K1 has a period close to 1 sol and is transient, unlike diurnally forced tides. Reliable identification of K1 therefore requires observations with high frequency resolution and/or long, continuous records. This has limited our understanding of K1 in the Martian atmosphere.

The aim of this study is to detect the K1 signal using surface pressure data from the InSight lander. This lander provides highly accurate (~50 mPa) surface pressure measurements with a sub-second temporal resolution spanning more than one Martian year, although it contains some data gaps. Lange et al. (2022) investigated long-term and interannual variations in Martian surface pressure, primarily using InSight observations and comparing them with measurements from Viking and the Mars Science Laboratory. Baroclinic waves (Banfield et al., 2020), thermal tides (Hernández et al., 2024), gravity waves (Hernández et al., 2025), and atmospheric turbulence (Spiga et al., 2021; Chatain et al., 2021) have already been analyzed in the InSight pressure data. The time resolution of the pressure sensor is sufficiently high to detect the high-order harmonics of thermal tides, including components with periods shorter than one hour (Hernández et al., 2024). In addition, InSight's near-equatorial location (4°N) is advantageous for detecting K1, as the theoretical latitudinal structure of K1's pressure component peaks at the equator. In the present study, singular spectrum analysis (SSA; Ghil et al., 2002) is employed to distinguish K1 from diurnal tides. SSA is commonly used to analyze non-stationary, quasi-periodic signals. Thus, this method enables us to distinguish between trends, quasi-periodic signal components, and noise. The analyzed period extends from the end of MY34 to the beginning of MY36, including the entire MY35. Our focus is on the seasonal variations in the amplitude and phase of K1, and its relation with dust storms.

In this paper, tidal harmonics with periods of 24 and 12 Mars hours are referred to as S1 and S2, respectively. Following Hernández et al. (2024), seasons are defined as 90° intervals in Ls centered on the equinoxes and solstices: $L_s$ = 315°–45°, 45°–135°, 135°–225°, and 225°–315° for the NH spring, summer, autumn, and winter, respectively.

## 2 Data and Methods

### 2.1 InSight surface pressure dataset

The K1 as well as S1 and S2 on Mars are investigated using the surface pressure timeseries acquired by the InSight lander, which provides exceptional precision (50 mPa) and extensive temporal coverage (Spiga et al., 2018; Banfield et al., 2019, 2020). Our analysis covers the period from $L_s = 304°$ in MY 34 to $L_s = 20°$ in MY36, corresponding to InSight sols 15–824. We used the processed InSight dataset produced by Hernández et al. (2024). Hernández et al. (2024) identified 99 intervals of continuous measurements, one of which had missing periods exceeding 70 s and assigned IDs ranging from ID#0 to ID#98 to these identified intervals. Gaps shorter than 70 s were filled using linear interpolation (see their supporting information for details). In the present study, SSA is applied separately to each interval ID; the procedure is described in the next section.

### 2.2 Singular spectrum analysis (SSA)

SSA provides a way to extract information from timeseries of limited-length that include observational noise (e.g., Ghil et al., 2002). SSA decomposes a timeseries to a trend, oscillatory components, and noise. An advantage of SSA is that the extracted trend does not have to be linear, and the periodic components can be non-stationary, allowing for amplitude and phase modulation. Broomhead and King (1986) and Sauer et al. (1991) therefore suggest that SSA is potentially efficient for exploring nonlinear dynamics. Unlike the forced diurnal tide S1, K1 is a normal mode that can be sporadically excited and affected by nonlinear processes, including resonance with DE1. Thus, K1 is likely to manifest as a quasi-periodic and non-stationary signal, for which SSA is well-suited. Previous studies of Martian surface pressure have used SSA to extract baroclinic wave signals (Collins et al., 1996) and to remove trends (Zurita-Zurita et al., 2022). However, no studies have been conducted to extract K1 using SSA.

In this study, SSA was applied to each interval timeseries that was longer than 10 sols. The first 30 reconstructed components, ordered by their respective explained variance, were retained. The dominant frequency of each reconstructed component was estimated using a Fourier transform. Since an oscillatory signal is represented by a pair of phase-shifted components, SSA usually identifies one oscillatory mode as two components with nearly equal explained variance and frequency. We therefore treated such a pair as a single periodic mode. K1 was detected in 17 out of 99 intervals. These intervals are summarized in table 1.

| ID | Trend (%) | S1 (%) | S2 (%) | Kelvin (%) |
|---|---|---|---|---|
| 7 | 92.1 | 3.65 | 1.47 | 0.760 |
| 8 | 94.0 | 2.80 | 1.38 | 0.097 |
| 9 | 94.9 | 2.29 | 1.11 | 0.084 |
| 10 | 95.2 | 2.23 | 1.07 | 0.028 |
| 12 | 95.7 | 1.86 | 1.08 | 0.070 |
| 12 | 95.8 | 1.86 | 0.87 | 0.120 |
| 13 | 96.7 | 2.03 | 0.33 | 0.028 |
| 37 | 96.9 | 1.86 | 0.37 | 0.028 |
| 67 | 94.3 | 2.71 | 1.17 | 0.160 |
| 71 | 93.6 | 2.15 | 1.02 | 0.099 |
| 73 | 95.1 | 2.48 | 1.27 | 0.035 |
| 75 | 95.6 | 2.15 | 0.94 | 0.072 |
| 76 | 95.6 | 1.99 | 0.89 | 0.069 |
| 77 | 95.2 | 1.90 | 0.96 | 0.086 |
| 87 | 95.7 | 1.94 | 0.92 | 0.036 |
| 95 | 96.1 | 1.58 | 0.84 | 0.043 |

**Table 1**. Contributions of the S1, S2, and K1 components to the total variance, as determined by singular spectrum analysis (SSA), for each interval (ID #) in which the Kelvin wave was detected.

Figure 1 shows the power spectral density (PSD) periodogram for interval ID#7, in which K1 was detected at the lowest SSA modes order among Interval IDs #0–99. Interval ID #07 corresponds to InSight sols 38–65, covering $L_s$~318°–$L_s$~335° in MY 34. The gray curve shows the PSD of the original timeseries and the red curve shows the PSD of the signal reconstructed from the first 30 SSA modes. Most variability with periods longer than 12 Mars hours (i.e., lower than two cycles per sol) is represented by these first 30 SSA modes. Variability at periods shorter than 12 Mars hours mainly corresponds to higher-harmonic tidal components, with no other distinct signals present. This result indicates that the reconstruction also successfully reduces noise as well. In interval ID#7, K1 is primarily identified in the 6th and 7$^{th}$ SSA modes, which are shown in magenta. It should be noted that for all intervals, the trend, S1, and S2 are extracted as the first, second–third, and fourth–fifth SSA modes, respectively. In contrast, the mode numbers corresponding to K1 vary from interval to interval, likely reflecting the transient nature of K1.

### 2.3 Mars Planetary Climate Model

We compare our results to the Laboratoire de Météorologie Dynamique Mars Global Climate Model (Forget et al., 1999), hereafter referred to as Mars planetary climate model (PCM) (Pottier et al., 2017). For this purpose, we ran specific simulations with a horizontal resolution of 1° × 1°, and outputs every 1/8 Martian hour, corresponding to 7.5 Martian minutes (usual model outputs are sampled at lower frequency, which would be insufficient for our purposes). Spiga et al. (2018) showed that Mars PCM realistically reproduces the surface pressure seasonal variations observed by InSight. The simulations were performed using dust scenarios for years MY34 and MY35 provided by Montabone et al. (2015; 2020). For a direct comparison with observations, our study only uses timeseries at a model cell that is closest to the actual InSight position for MY34 and MY35.

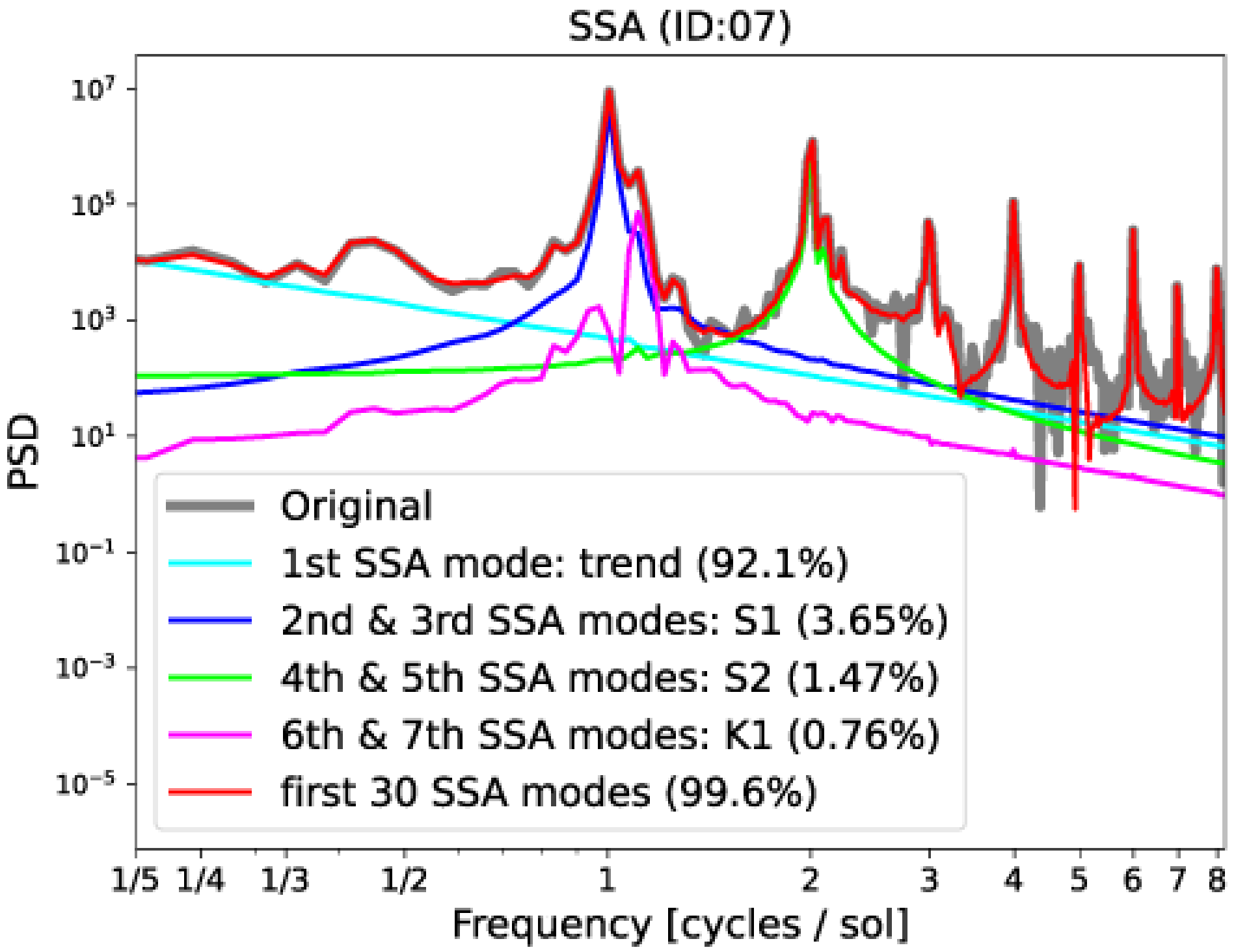

**Figure 1**. Periodogram of Power Spectral Density (PSD) as a function of frequency (cycles per sol) for a ~29-sol interval with continuous coverage (ID#07). The thick gray curve shows the PSD of the original data, and the red curve shows the PSD of the reconstructed signal using the first 30 SSA modes.

## 3 Results from the InSight data

### 3.1 Seasonal variations of K1 amplitudes

Figure 2 shows the seasonal variations of amplitudes for (a) the trend component, (b) the S1 diurnal signal, (c) the S2 semi-diurnal signal, and (d) the K1 quasi-diurnal signal for the time period from $L_s$ ~ 315° in MY34 to $L_s$ ~ 20° in MY36. The trend component clearly shows the annual cycle associated with CO2 condensation and sublimation on polar caps and seasonal changes of global-scale atmospheric circulation (Spiga et al., 2018; Lange et al., 2022). The magnitude of the trend component is almost comparable to the pressure at the InSight landing site from the Mars Climate Database (MCD) version 5 (Millour et al., 2024): The pressure minimum of about 600 Pa is observed at around $L_s$ ~150°, while the pressure maximum of about 800 Pa occurs at around $L_s$ ~260°, just before the NH winter solstice (Figure 2a). The S1 component ranges from ~10 to ~20 Pa throughout the Martian year. During InSight sols of

∼40– ∼50 corresponding to the C-storm in MY34 (C34), however, the amplitude of S1 increases abruptly up to ∼39 Pa (Figure 2b). The S2 amplitude exhibits a similar sudden increase to S1 with its maximum of ∼13 Pa during the C34 storm, while it usually ranges from ∼5 to ∼10 Pa throughout the Mars year (Figure 2c). The significant K1 components were mainly detected during the following three time periods: First, during the C34 storm, the magnitude of K1 component peaks at around 8 Pa, and then decreases rapidly, as seen for S1 and S2. Second, during the latter half of NH spring ($L_s$ ∼ 0°–45°) the K1 amplitude reaches up to ∼1 Pa. Third, during NH autumn ($L_s$ ∼ 150°–225°), the K1 amplitude ranges from ∼1 to ∼2 Pa. The absence of significant K1 detections during NH summer ($L_s$ ∼ 45°–135°) are possibly explained as follows: Due to the globally lower atmospheric temperatures during NH summer, and as mentioned in the introduction, the theoretical frequency of K1 is quite close to the diurnal frequency (Wilson & Hamilton, 1996), which makes it more difficult to separate K1 from S1 even by SSA. In addition, the InSight pressure record during this NH summer period may be too short for reliable K1 detection. During the remaining part of NH winter ($L_s$ ∼ 225°–315°), K1 like signals that have quasi-diurnal frequencies were identified in several intervals; however, the explained variances of the identified signals are smaller than 0.1%. This low value indicates that the separation between signal and noise can be ambiguous (Ghil et al., 2002).

Given these results, in the following subsection, we focus on the interval ID#07 during the MY34 C storm, when K1 has the largest explained variance of 0.76% of total. Note that this K1 variance is approximately 1/2 of that of S2 and 1/5 of that of S1, respectively, whereas in most other interval IDs, it is less than one-twentieth that of S1.

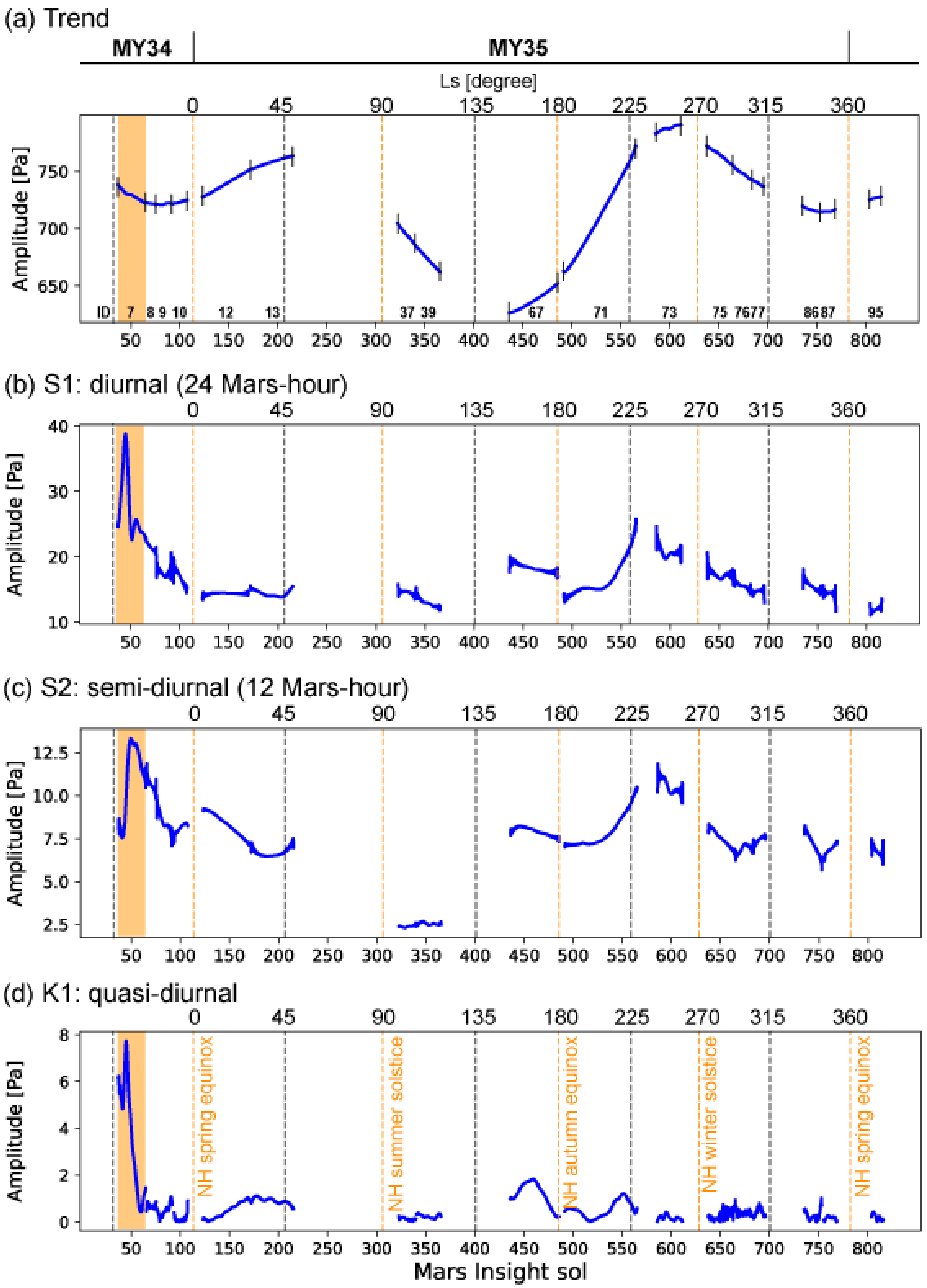

(a) Trend
MY34
MY35
Ls [degree]
0
45
90
135
180
225
270
315
360
Amplitude [Pa]
ID
7
8 9 10
12
13
37 39
67
71
73
75 7677
8687
95
(b) S1: diurnal (24 Mars-hour)
(c) S2: semi-diurnal (12 Mars-hour)
(d) K1: quasi-diurnal
NH spring equinox
NH summer solstice
NH autumn equinox
NH winter solstice
NH spring equinox
Mars Insight sol

**Figure 2**. Timeseries of surface pressure amplitudes for all continuous intervals in which K1 is detected: (a) the trend, (b) S1, (c) S2, and (d) K1. The short vertical lines in the timeseries in (a) mark the start and end of each interval with the ID label at the bottom of Figure 2a. Orange boxes indicate the interval ID#07 during the C-storm in MY34, which this study focuses on.

### 3.2 Phase of K1 and S1 during C storm in MY34

Figure 3 provides the sol and local true solar time cross sections of the K1, S1, and S2 components of the SSA decomposition during Mars InSight sols 38–65 in the C34 storm period ($L_s$ =∼318–∼334). Since thermal tides are atmospheric responses to solar insolation, their phases are fixed with respect to the local true solar time. The wave period of K1 is slightly shorter than 24 Mars hours. Therefore, as the sol progresses, the K1 phase appears to shift to earlier local times, as shown in Figure 3a. The K1 amplitude reaches a maximum of ∼8 Pa around sols 44–47 after which it gradually decreases. The S1 amplitude increases during the ∼7 sols following sol 38 and reaches ∼40 Pa during sols 44–48, which is about twice the value observed on sol 38. The S1 phase exhibits interesting behavior: when the amplitude is near its maximum, the phase shifts to earlier local times. At sol 50, the S1 phase is ∼3 Mars hours earlier than at sol 40, returning to its original phase by sol 54. During this S1 phase shift toward earlier local times, K1 is in phase with S1. However, during the return to the original phase, K1 is approximately out of phase with S1. In contrast, the S2 amplitude reaches its maximum during sols 50–56. This peak occurs ∼5 sols later than those of K1 and S1. In general, it is considered that S2 responds strongly to the amount of airborne dust [e.g., Guzewich et al., 2016; Viúdez-Moreiras et al., 2020]. Therefore, it appears that the amplification of K1 and S1 serves as a precursor to dust storms, as proposed in previous studies. In the next section, a quantitative analysis of the phase relation between K1 and S1 is made.

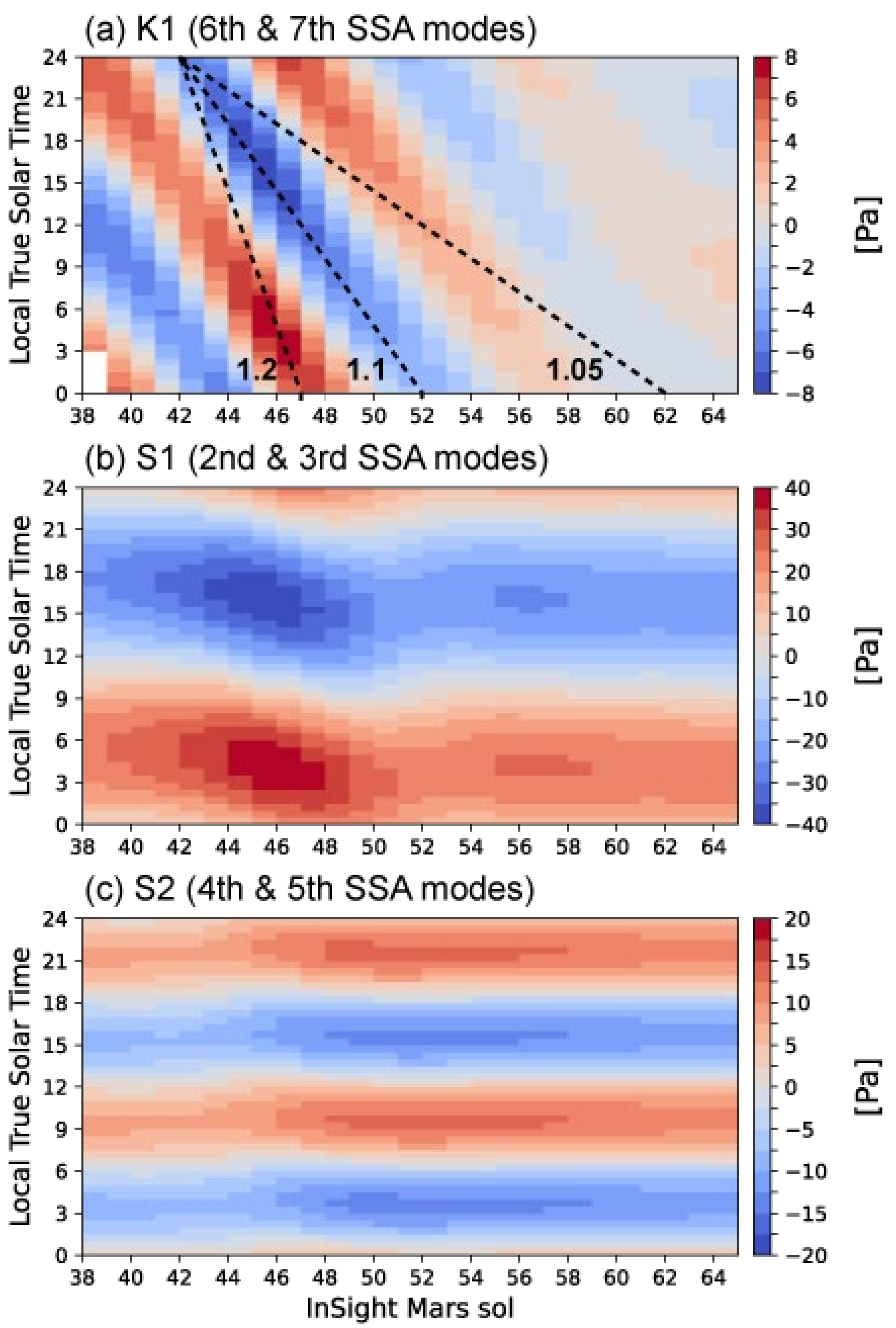
(a) K1 (6th & 7th SSA modes)
Local True Solar Time
1.2
1.1
1.05
[Pa]
(b) S1 (2nd & 3rd SSA modes)
Local True Solar Time
[Pa]
(c) S2 (4th & 5th SSA modes)
Local True Solar Time
InSight Mars sol
[Pa]

**Figure 3**. The InSight Mars sol and local true solar time sections of phase and amplitude for (a) K1, (b) S1, and (c) S2. The black dotted lines in Figure 3a indicate reference frequencies for K1; from left to right, they are 1.20, 1.10, and 1.05 cycle/sol.

### 3.3 The relation between K1 and S1 amplitudes with zonally non-uniform dust distribution

Figure 4a shows the time evolution of the phase relation between K1 and S1 during the C34 storm, as well as their respective amplitudes, which are plotted as red and blue curves on the right y-axis. The red and blue dots show the estimated periods of K1 and S1 referenced to the left y-axis, which are defined as the time intervals between successive local maximum phases and referred to as the apparent period. The dashed blue and red horizontal lines indicate the reference periods of 24 Martian hours for S1 and approximately 21.8 Martian hours (∼1.1 cycles per sol) for K1, respectively. The reference frequency of K1 (∼1.1 cycles per sol) corresponds to the peak in the PSD shown in Figure 1. Although the S1 component is a diurnal thermal tide forced by periodic solar heating, the apparent S1 period (blue dots) exhibits measurable variations. Starting at sol 42, the S1 period decreases below 24 Mars hours and reaches a minimum of ∼23.5 Mars hours at sol 45. The apparent S1 period then increases gradually over the next five sols, reaching ∼24.5 Mars hours at sol 51. It subsequently decreases back toward 24 Mars hours and, after sol 54, remains stably around 24 Mars hours. The apparent K1 period varies between ∼21 and ∼22 Mars hours until about sol 54. After that, the K1 period largely scatters, owing to much smaller amplitudes. As shown in Figure 3, S1 and K1 are in phase when their amplitudes reach a maximum of ∼40 Pa (S1) and ∼8 Pa (K1), respectively. During this interval, the apparent S1 period is shorter than 24 Mars hours. This near phase-locking during the shortened S1 period is consistent with the resonant interaction between the eastward nonmigrating diurnal tide DE1 and the quasi-diurnal Kelvin mode K1. After sol ∼49, S1 and K1 are out of phase. This change in the difference between the S1 and K1 phases is likely related to the change in the apparent S1 period, which is longer than 24 Mars hours during this interval.

Next, we analyzed the total column dust optical depth, CDOD, from the Mars Climate Database (MCD; Montabone et al., 2020; Millour et al., 2024). Figure 4b presents the timeseries of longitudinal-mean and latitudinal area-weighted mean CDOD by a thick dark-red curve. The zonal wavenumber-1 and wavenumber-2 components of CDOD around the equator, i.e., 20°S–20°N, are also calculated to examine the zonal non-uniformity of dust distribution, as respectively shown by the hatched blue and orange curves. Both the zonal wavenumber-1 and wavenumber-2 components peak around sols 46–47, followed by a decline showing global dust spread. In fact, a few sols after the zonal wavenumber-1 and wavenumber-2 component maxima, the global-mean dust peaks around sols 48–51 and gradually decreases thereafter. When the zonal wavenumber-1 and wavenumber-2 components are large, in other words, when dust is zonally non-uniformly distributed, the S1 amplitude is large and the apparent S1 period is shorter than 24 Mars hours.

Using a Mars GCM with and without topography, Wilson and Hamilton (1996) suggested that DE1 is generated through nonlinear interaction between the diurnal westward migrating tide (DW1) and the zonal wavenumber-2 component of equatorial topography. This mechanism can be illustrated by the following equation:

$$\cos(2\lambda)\cos(\Omega t_{\mathrm{LT}}) = \frac{1}{2}[\cos(\Omega t_{\mathrm{UT}} + 3\lambda) + \cos(\Omega t_{\mathrm{UT}} - \lambda)], \quad (1)$$

where $\lambda$ is longitude, $\Omega$ is diurnal frequency, $t_{\mathrm{LT}}$ is local true solar time, and $t_{\mathrm{UT}}$ is universal time referenced to a fixed meridian. Note that the $t_{\mathrm{LT}}$ and $t_{\mathrm{UT}}$ are related by

$$t_{\mathrm{LT}} = t_{\mathrm{UT}} + \lambda/\Omega. \quad (2)$$

The factor $\cos(2\lambda)$ represents zonal wavenumber-2 inhomogeneity at the lower boundary like topography, and the $\cos(\Omega t_{\mathrm{LT}})$ is Sun-synchronous forcing. Thus, this interaction yields two diurnal components $\cos(\Omega t_{\mathrm{UT}} + 3\lambda)$ and $\cos(\Omega t_{\mathrm{UT}} - \lambda)$. The latter, $\cos(\Omega t_{\mathrm{UT}} - \lambda)$, is DE1.

During the early stage of the dust storm, the dust loading can be zonally non-uniform, which provides strong zonally asymmetric thermal forcing similar to the above-mentioned topographic forcing but transient in nature. It is inferred that this transient forcing enhances DE1 and may simultaneously excite K1. When DE1 and K1 are in phase, both are amplified. This amplification is probably related to their similar wave periods of 1 sol and their same meridional structures, which make resonant interaction more likely. Figure 3 of Viúdez-Moreiras et al. (2020) shows four snapshots of the longitude–latitude distribution of column dust optical depth during the C34 storm, corresponding to InSight sols 41, 44, 48, and 52. Their Figures 3b–3e show that the dust distribution expands longitudinally as the event develops.

Recent studies have shown that both rapid meridional dust transport on subsol timescales, occurring primarily in the Southern Hemisphere, and zonal dust transport in the equatorial region are associated with winds driven by thermal tides (Wu et al., 2020, 2025; Cheng et al., 2025; Bertrand et al., 2020). Wu et al. (2020, 2025) and Cheng et al. (2025) diagnosed these tidal winds from temperature fields observed by the Mars Climate Sounder (MCS) using classical tidal theory (Chapman and Lindzen, 1969). Such rapid, tidally driven dust motion within a sol has been referred to as a "dust tide." More recently, Toigo et al. (2026) analyzed observations from the Emirates Mars Mission Emirates Mars Infrared Spectrometer (EMM/EMIRS), whose broad local-time coverage and relatively high spatiotemporal resolution revealed the development of a dust storm over only a few sols. Using MarsWRF simulations, Toigo et al. (2026) investigated how tidal winds affect the expansion of the dust storm. Their results suggest that dust storms can expand not only meridionally but also zonally, and that the direction and extent of this expansion may depend on the local spatial structure of the thermal tides. In light of these findings, our results suggest a possible dynamical mechanism for the zonal expansion of large-scale dust storms. The theoretical meridional structures of both DE1 and K1 are characterized by large amplitudes of zonal wind components around the equator. Thus, the zonal winds associated with DE1 and K1 may have contributed to the longitudinal transport of dust during the development of the large-scale dust storm.

In contrast to the time evolution of S1 and K1 amplitudes, the S2 amplitude increases later and peaks at about sols 50–56 (Figure 3c). This later enhancement of S2 is in line with the temporal evolution of the globally averaged dust opacity which is maximized around sols 48–49.

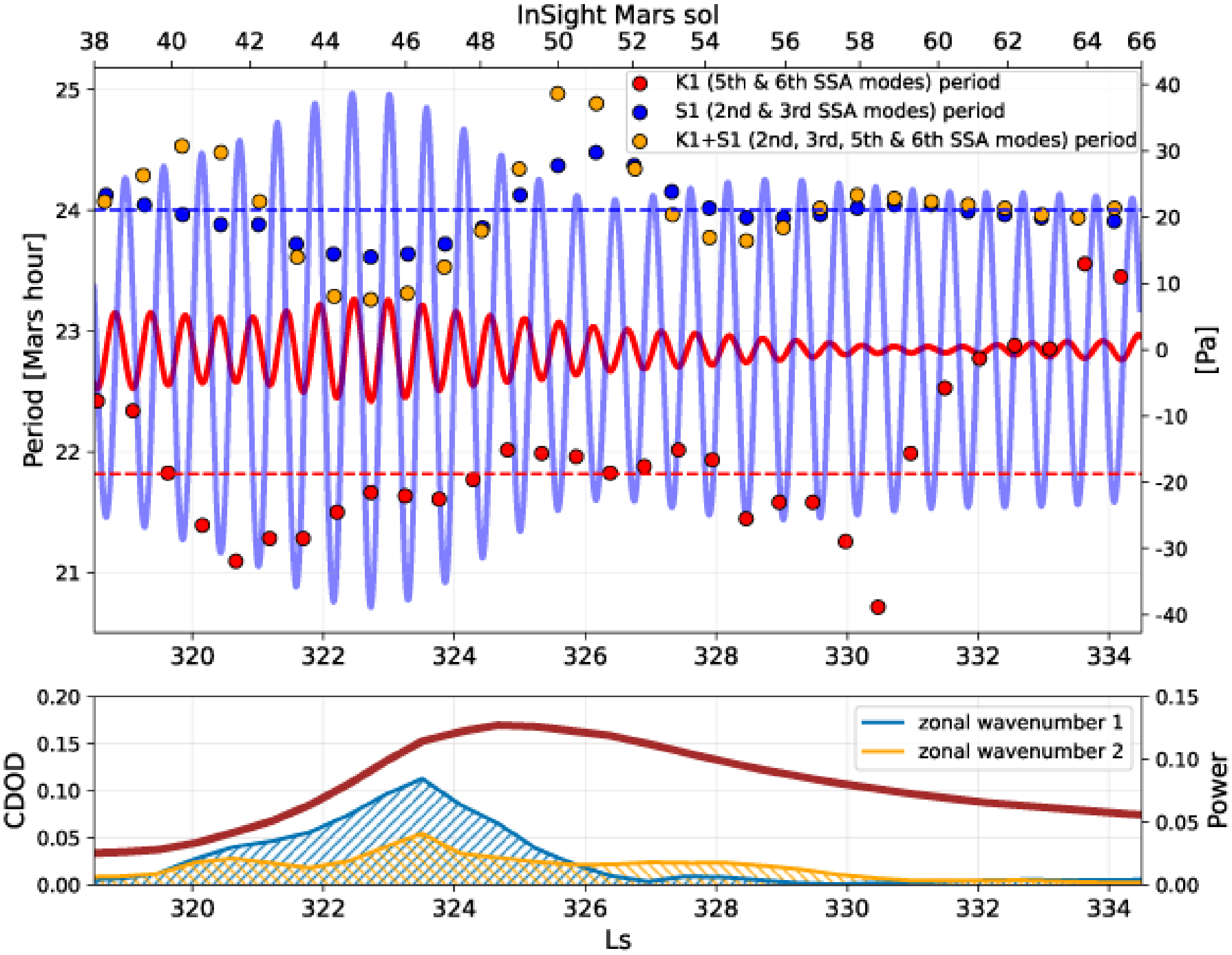

**Figure 4**. (a) Timeseries of the S1 signal (blue curve) and K1 signal (red curve). Blue and red dots mark the estimated periods of S1 and K1, computed from the time interval between successive local maxima. Orange dot denotes the apparent periods of the signal resulting from the interference of S1 and K1. (b) Timeseries of total dust total Column dust optical depth CDOD from Mars Climate Database (Millour et al., 2024) during the MY34 C-storm. The thick dark-red line shows the zonal-mean and latitudinally area-weighted mean, namely, global-mean dust optical depth. The hatched blue and red curves show the zonal wavenumber-1 and wavenumber-2 components of dust opacity, respectively.

## 4 Results from Mars PCM simulations

Mars PCM simulations were performed under the MY34 and MY35 dust scenarios (Figures 5a and 5b). Using the simulated surface pressure at the model grid point nearest to the InSight landing site, we divided the data into six intervals: sols 0–110, 111–220, 221–330, 331–440, 441–550, and 551–668. The interval lengths were determined empirically through sensitivity tests to ensure stable and reliable detection of the K1 signal. SSA was then applied separately to each interval. Figure 5 shows the seasonal variations in the K1 amplitude. In both MY34 and MY35, a clear K1 signal was detected during sols 551–668 ($L_s$ ~290–360), as shown by the orange boxes. These periods correspond to the seasonal C-storm that occurs in each Mars year. During the C34 storm, K1 was identified as the 8th and 9th SSA modes, following S1 (2nd and 3rd SSA modes), S2 (4th and 5th SSA modes), and S4 (6th and 7th SSA modes). During C35, K1 was identified as the 6th and 7th SSA modes, following S1 (2nd and 3rd SSA modes) and S2 (4th and 5th SSA modes). As a reminder, the SSA modes are ordered by their explained variance. It should be noted that K1 was detected in the InSight data during the C34 storm, but not during the C35, probably because of a large InSight data gap during the C35 period (Figure 3).

Figures 5a2–a4 and 5b2–b4 show sol–local time cross sections of the K1, S1, and S2 components during sols 550–668, including the C34 storm and C35 storm periods. The estimated K1 frequency is 1.02 cycle/sol for both MY34 and MY35, which is very close to the diurnal period. The time evolution of the K1 amplitude seen in the Mars PCM results resembles the InSight results discussed in Section 3.2: the K1 amplitude reaches its maximum around Sol 610 in MY34 (Figure 5a2) and around Sol 600 in MY35 (Figure 5b2), when K1 and S1 are in phase. Thereafter, the K1 amplitude decreases rapidly. However, unlike the InSight results presented in Section 3.3, the Mars PCM does not clearly show a shortening of the apparent S1 period, that is, a phase shift toward earlier local times, when the phases of K1 and S1 match. This difference between Mars PCM simulation and InSight observations may be explained by the fact that the estimated K1 frequency in the Mars PCM (1.02 cycles per sol) is much closer to the diurnal frequency than estimated frequency from the InSight data (1.1 cycles per sol). Thus, the K1–S1 interaction in the Mars PCM occurs closer to the diurnal frequency than the real atmosphere. As a result, the apparent phase shift or shortening of S1 period may be less evident in the Mars PCM. The reason why the K1 frequency is lower in the Mars PCM than in the InSight observations remains unclear, and a detailed investigation is beyond the scope of the present study. Wilson and Hamilton (1996) suggested that the theoretical period of K1 is sensitive to the global-mean atmospheric temperature. One possible explanation is therefore

that the lower K1 frequency in the Mars PCM reflects a lower simulated global-mean atmospheric temperature during the C34 storm than in the real atmosphere.

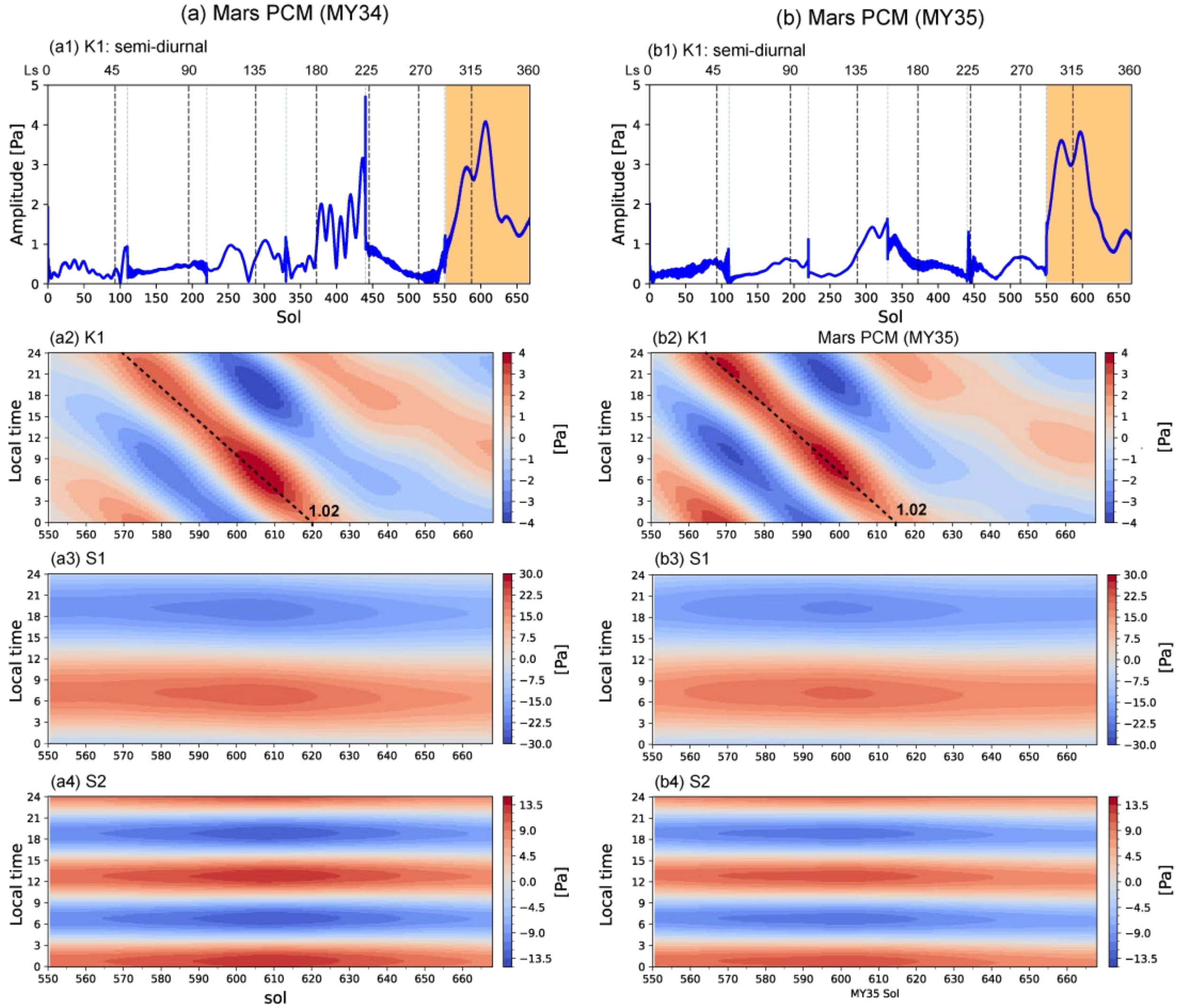

**Figure 5**. Timeseries of surface pressure amplitudes of K1 in Mars PCM under (a1) the MY34 dust scenario and (b1) the MY35 dust scenario. Orange boxes indicate the periods when the K1 signal was most clearly detected, which correspond to the C-storm in MY34 and MY35, respectively. (a2–a4): Sol–local time sections of phase and amplitude of K1 (a2), S1 (a3), and S2 (a4) during sols 550–668 in MY34. (b2–b4): Same as Figures a2–a4 but for MY35. The black dotted lines in Figures a2 and b2 indicate reference frequencies for K1, 1.02 cycle/sol.

## 5 Summary and concluding remarks

In this study, the quasi-diurnal Kelvin wave K1 was successfully detected in the Martian atmosphere by applying SSA to the InSight surface pressure timeseries. The dataset spans more than one Martian year, including the MY34 C-storm period and the entirety of MY35 with high temporal resolution and continuity. The combination of SSA and this extensive InSight dataset enabled the direct extraction of K1, in contrast to previous studies that inferred K1 indirectly from transient phase modulations of the diurnal signal S1 (Haberle et al., 2014; Viúdez-Moreiras et al., 2020; Zurita-Zurita et al., 2022).

Seasonal variations of K1 amplitudes are summarized as follows: K1 is rarely detected during NH summer. This is likely because the globally lower temperature during this season brings its theoretical frequency close to the diurnal frequency, which makes it difficult to distinguish K1 from diurnal tides (Wilson & Hamilton, 1996). During NH spring and autumn, relatively clear K1 signals are detected, with amplitudes of about 1–2 Pa (section 3.1). During the MY34 C-storm, K1 amplitudes reached up to about 8 Pa, approaching the magnitude of the S1 and S2 amplitudes, which reached about 39 Pa and 13 Pa, respectively.

The phase relation between S1 and K1 during the MY34 C-storm is examined in detail using InSight data. When K1 and S1 become in phase at $L_s$ 332°–334° in MY34, both amplitudes reach their maxima. During this time period, the apparent S1 period shortens to about 23.5 Mars hours. The observed phase matching and the shortening of the wave period suggest a resonant interaction between K1 and the eastward, non-migrating diurnal tide (DE1). Moreover, the temporal evolutions of the S1 and K1 amplitudes appear to accord with that of the zonally non-uniform distribution of dust. Recent studies have reported that thermally forced diurnal tidal winds drive intense diurnal dust motion (e.g., Wu et al., 2020; Toigo et al., 2026). Given these findings, together with our observational results, we consider the dust storm spreading mechanism to be as follows: The localized dust loading around the equator in the early stage of the dust storm likely acts as transient zonally asymmetric forcing that enhances DE1 and simultaneously excite K1. The zonal motion of dust is coupled with the zonal winds associated with DE1 and K1 as these waves are amplified through resonant interaction. Dust gradually spreads as the DE1 and K1 propagate.

The Mars PCM simulations under the MY34 and MY35 dust scenarios were performed and analyzed using the same SSA method. The result shows that there are clear K1 signals during the C-storm periods of both years. K1 is commonly enhanced when it is approximately in phase with S1, and then its amplitude decreases rapidly, resembling the temporal evolution observed by InSight during the C34 storm. However, the Mars PCM does not clearly reproduce the apparent shortening of the S1 period observed by InSight when K1 and S1 are in phase. This difference between the model and observation is attributable to the difference in the K1 frequency: the estimated K1 frequency in the Mars PCM is approximately 1.02 cycles per sol, which is much closer to the diurnal frequency than the value of approximately 1.1 cycles per sol estimated from the InSight data. Thus, the interaction or resonance between K1 and DE1 seen in the Mars PCM occurs at almost 1 cycle/sol.

In future work, a comprehensive investigation of the three-dimensional structure of K1 using Martian general circulation models and/or reanalysis datasets will be needed. Full longitudinal coverage would enables distinction between eastward and westward propagation. Wide latitudinal and vertical coverage would allow us to examine the meridional and vertical structures of K1, including its expected equatorial amplitude maximum. These analyses would help elucidate the excitation mechanism of K1 associated with zonally nonuniform dust loading and improve our understanding of the expansion processes of Martian dust storms.

**Acknowledgments**

We greatly appreciate constructive comments by Kevin Hamilton, and Ehouarn Millour for providing the Mars PCM runs.

**Open Research**

The continuous time-series dataset used for this study is available from Hernández-Bernal and Spiga (2023). The InSight data analyzed in this study are publicly accessible through the Planetary Data System (PDS; Banfield, 2020).

**Conflict of Interest Disclosure**

The authors have no conflicts of interest to disclose.